\input{aipcheck}

\documentclass[
    ,final            
  ]
  {aipproc}

\layoutstyle{8x11double}

\begin{document}

\def\bea{\begin{eqnarray}}
\def\eea{\end{eqnarray}}
\def\bec{\begin{center}}
\def\ec{\end{center}}
\def\pC{\tilde{\chi}^+}
\def\nC{\tilde{\chi}^-}
\def\pnC{\tilde{\chi}^{\pm}}
\def\Ne{\tilde{\chi}^0}
\def\snu{\tilde{\nu}}
\def\tN{\tilde N}
\def\ler{\lesssim}
\def\gtr{\gtrsim}
\def\beq{\begin{equation}}
\def\eeq{\end{equation}}
\def\haf{\frac{1}{2}}
\def\lpp{\lambda''}
\def\ccg{\cal G}
\def\slash#1{#1\!\!\!\!\!/}
\def\rpv{\slash{R_p}}
\def\ler{\lesssim}
\def\gtr{\gtrsim}
\def\pslash{p\hspace{-2.0mm}/}
\def\qslash{q\hspace{-2.0mm}/}
\def\p{\partial}
\def\f{\frac}

\newcommand{\imag}{\Im {\rm m}}
\newcommand{\real}{\Re {\rm e}}

\def\f#1#2{\frac{#1}{#2}}
\def\p{\partial}
\def\th{\theta}
\def\si{\sigma}
\def\Si{\Sigma}
\def\phivis{\Phi_{\rm vis}}
\def\phibulk{\Phi_{\rm bulk}}
\def\pr{\prime}
\def\l{\left}
\def\r{\right}

\title{Moduli stabilization and the pattern of sparticle spectra}


\author{Kiwoon Choi}{
  address={Department of Physics, Korea Advanced Institute of Science
and Technology, Daejeon 305-701, Korea} }



\begin{abstract}
We discuss the  pattern of low energy sparticle spectra which
appears  in some class of
 moduli stabilization scenario.
In case that light moduli are stabilized by non-perturbative effects
encoded in the superpotential and a phenomenologically viable de
Sitter vacuum is obtained by a sequestered supersymmetry breaking
sector,
the anomaly-mediated soft terms become comparable to the
moduli-mediated ones, leading to a quite distinctive pattern of low
energy spacticle masses dubbed the mirage mediation pattern. We also
discuss low energy sparticle masses in more general mixed-mediation
scenario which includes a comparable size of gauge mediation in
addition to the moduli and anomaly mediations.
\end{abstract}

\maketitle


\section{introduction}

Low energy supersymmetry (SUSY) is one of the prime candidates for
physics beyond the standard model at TeV scale \cite{nilles}. One of
the key questions on low energy SUSY is the origin of soft SUSY
breaking terms of the visible gauge/matter superfields in the low
energy effective lagrangian \cite{kane}. Most of the
phenomenological aspects of low energy SUSY are determined by those
soft terms which are presumed to be induced by the auxiliary
components of some messenger fields. In string theory, moduli fields
including the string dilaton are plausible candidates for the
messenger of SUSY breaking \cite{modulus}. In addition to  string
moduli, the 4-dimensional supergravity (SUGRA) multiplet provides a
model-independent source of SUSY breaking, i.e. the anomaly
mediation \cite{anomaly}, which induces a soft mass
 $m_{\rm soft}\sim m_{3/2}/8\pi^2$.

To identify the dominant source of soft terms, one needs to compute
the relative ratios between different auxiliary components including
the auxiliary component of the 4D SUGRA multiplet. This requires an
understanding of how the messenger moduli are stabilized at a nearly
4D Poincare invariant vacuum.
In this talk, we discuss the  pattern of low energy sparticle
spectra which appears  in some class string compactifications which
realize the low energy SUSY at TeV scale while stabilizing all
moduli \cite{choi1,choi2}.

\section{4d effective sugra with sequestered susy breaking}

 Our theoretical framework is an effective SUGRA of
string compactification with a sequestered SUSY breaking sector. To
be specific, we will be focusing on KKLT-type compactification with
an warped throat which is produced by stringy flux \cite{gkp,kklt}.
The internal space of KKLT-type compactification consists of a bulk
space which might be approximately a Calabi-Yau (CY) manifold, and a
highly warped throat attached at CY with SUSY-breaking brane
stabilized at its IR end. In such geometry, the bulk CY  can be
identified as the UV end of throat. To realize the high scale gauge
coupling unification, the visible gauge and matter fields are
assumed to live on $D$ branes stabilized within the bulk CY. In the
following, we will assume that the SUSY breaking at the IR end of
throat is provided by an anti-brane \cite{kklt}, which might be the
simplest way to realize $N=1$ SUSY breaking at a meta-stable vacuum
in string theory. Although we are taking the SUSY breaking by
anti-brane for simplicity, the resulting sparticle spectra in the
visible sector are independent of this particular choice, and valid
even for generic type of SUSY breaking at the IR end of throat
\cite{choi2,choi3}.

The 4D effective theory of the KKLT-type compactification includes
the UV superfields
$\Phi_{UV}=\{T, U, \Sigma\}$ and $V^a,Q^i$, where $T$ and $U$ are
the K\"ahler and complex structure moduli of the bulk CY, $V^a$ and
$Q^i$ are the gauge and matter superfields confined on the visible
sector $D$ branes, and $\Sigma$ denotes the open string moduli on
those $D$ branes at the UV side. There are also 4D fields localized
at the IR end of throat, e.g. $\Phi_{IR}=\{Z, \xi^\alpha\}$, where
$Z$ is the throat (complex structure) modulus superfield
parameterizing the size of 3-cycle at the IR end, and $\xi^\alpha$
is the Goldstino fermion living on the SUSY breaking brane.
If SUSY is spontaneously broken by  a chiral superfield $Y$ on SUSY
breaking brane at the IR end, the Goldstino corresponds to the
fermion component of $Y$: \bea
Y=Y_0+\xi^\alpha\theta_\alpha+F^Y\theta^\alpha\theta_\alpha,\eea
where $\langle F^Y\rangle\equiv M_{\rm SUSY}^2$ sets the scale of
SUSY breaking. After integrating out $Y_0$ and $F^Y$,  SUSY appears
to be non-linearly realized. As is well known, low energy effective
action with non-linearly realized SUSY can be written on the $N=1$
superspace with the Goldstino superfield: \bea
\Lambda^\alpha=\frac{1}{M_{\rm
SUSY}^2}\xi^\alpha+\theta^\alpha+...,\eea where
the ellipses stand for the Goldstino-dependent higher order terms in
the $\theta$-expansion. If SUSY is explicitly broken by an
anti-brane as in the original KKLT proposal \cite{kklt}, there is no
degree of freedom corresponding to the $N=1$ superpartner of
$\xi^\alpha$. However still the low energy dynamics can be described
by an effective action on $N=1$ superspace with the Goldstino field
as in the low energy limit of spontaneously broken SUSY
\cite{choi1}.

 In addition to the above UV and IR
fields, there is of course the 4D SUGRA multiplet which is
quasi-localized in the bulk CY, and also the string dilaton
superfield $S$ whose wavefunction is approximately a constant over
the whole internal space.

Generic 4D effective SUGRA action can be written as
\bea
\label{4daction1}&& \int d^4x\sqrt{g}\left[\int d^4\theta \,CC^*
\left\{-3\exp \left(-\frac{K}{3}\right)\,\right\} \right.\nonumber
\\&&+\,\left.\left\{\,\int d^2\theta \,\left( \frac{1}{4}f_a
W^{a\alpha}W^a_\alpha+C^3W\right)+{\rm h.c}\,\right\}\,\right]\eea
where $g_{\mu\nu}$ is the 4D metric in the superconformal frame,
$C=C_0+F^C\theta^2$ is the 4D SUGRA compensator, $K$ is the K\"ahler
 potential, and
 $f_a=T+lS$ ($l=$ rational number)
  are holomorphic gauge kinetic functions which are assumed to
be universal to accommodate the high scale gauge coupling
unification.
The UV and IR fields are geometrically separated by warped throat,
thus are {\it sequestered} from each other in $e^{-K/3}$:
\bea -3\exp\left(-\frac{K}{3}\right)&=&\Gamma_{UV}+\Gamma_{IR}, \eea
where \bea\Gamma_{UV}&=&
\Gamma_{UV}^{(0)}(S+S^*,\Phi_{UV},\Phi^*_{UV}) \nonumber \\
&+&{\cal Y}_i(S+S^*,\Phi_{UV},\Phi^*_{UV})Q^{i*}Q^i, \nonumber
\\\Gamma_{IR}
&=& \Gamma_{IR}^{(0)}(S+S^*,Z,Z^*)\nonumber \\
&+&\left(
\frac{C^{*2}}{C}\Lambda^2\Gamma_{IR}^{(1)}(S+S^*,Z,Z^*)+{\rm
h.c}\right)
\nonumber \\
&+&CC^*\Lambda^2\Lambda^{*2}\Gamma_{IR}^{(2)}(S,S^*,Z,Z^*)+..., \eea
where $\Phi_{UV}=\{T, U, \Sigma\}$, and  $\Gamma_{IR}$ is expanded
in powers of the Goldstino superfield $\Lambda^\alpha$ and the
superspace derivatives
$D_A=\{\partial_\mu,D_\alpha,\bar{D}_{\dot{\alpha}}\}$.
 The above
effective action is written on flat superspace background and the
SUSY-breaking auxiliary component of the 4D SUGRA multiplet is
encoded in the $F$-component of the compensator $C$. In the
superconformal gauge in which $C=C_0+F^C\theta^2$, the 4D action is
invariant under the rigid Weyl transformation under which \bea
\label{weyl} && C\rightarrow e^{-2\sigma}C,\quad
g^C_{\mu\nu}\rightarrow e^{2(\sigma+\sigma^*)}g^C_{\mu\nu},\quad
\nonumber \\
&& \theta^\alpha \rightarrow
e^{-\sigma+2\sigma^*}\theta^\alpha,\quad \Lambda^\alpha \rightarrow
e^{-\sigma+2\sigma^*}\Lambda^\alpha, \eea where $\sigma$ is a
complex constant, and this determines for instance the
$C$-dependence of $\Gamma_{IR}$.

The effective superpotential of KKLT compactification contains three
pieces: \bea W&=&W_{\rm flux}+W_{\rm np}+W_{\rm Yukawa},\eea where
the flux-induced  $W_{\rm flux}$ stabilizing $S,U,Z,\Sigma$ includes
the Gukov-Vafa-Witten superpotential  $W_{GVW}=\int (F_3-4\pi iS
H_3)\wedge \Omega$, where $\Omega$ is the holomorphic $(3,0)$ form
of the underlying CY space,
$W_{\rm np}$ is a non-perturbative superpotential stabilizing $T$,
and finally $W_{\rm Yukawa}$ denotes the Yukawa couplings of the
visible matter fields. Generically, each piece takes the form:
 \bea W_{\rm flux} &=&
\Big({\cal F}(U,\Sigma)+\frac{N_{RR}}{2\pi i}Z\ln Z+{\cal
O}(Z^2)\Big)
\nonumber \\
&&-\,4\pi
iS\Big({\cal H}(U,\Sigma)+N_{NS}Z+{\cal O}(Z^2)\Big),\nonumber \\
W_{\rm np}&=&{\cal A}(U,\Sigma)e^{-8\pi^2 (k_1T+l_1S)},
\nonumber \\
W_{\rm Yukawa}&=& \frac{1}{6}\lambda_{ijk}(U,\Sigma)Q^iQ^jQ^k,\eea
where $k_1,l_1$ are rational numbers,  $N_{RR},N_{NS}$ are integers
defined as $N_{RR}=\int_{\Sigma} F_3,
N_{NS}=-\int_{\tilde{\Sigma}}H_3$, where $\Sigma$ is the 3-cycle
collapsing  along the throat, $\tilde{\Sigma}$ is its dual 3-cycle,
and $F_3$ and $H_3$ are the RR and NS-NS 3-forms, respectively.
 Here, we assumed that the axionic shift symmetry
of $T$, i.e. $T\rightarrow T+$ imaginary constant, is preserved by
$W_{\rm flux}$ and $W_{\rm Yukawa}$, but is broken  by $W_{\rm np}$.
Note that $Z$ is defined as $\int_{\Sigma}\Omega=Z$, and then
$\int_{\tilde{\Sigma}}\Omega=\frac{1}{2\pi i}Z\ln Z
+\mbox{holomorphic}$ \cite{gkp}.

The  above 4D effective action of KKLT-type compactification
involves many model-dependent functions of moduli,
which are difficult to be computed for realistic compactification.
Fortunately, the visible sector soft terms can be determined by only
a few information on the compactification, e.g. the rational
parameters $l,k_1,l_1$ in $f_a$ and $W_{\rm np}$ and the modular
weights which would determine the $T$-dependence of ${\cal Y}_i$,
which can be easily computed or parameterized in a simple manner.
 In particular, soft terms are {\it practically independent} of the detailed forms of
  $\Gamma^{(0)}_{UV}$, $\Gamma_{IR}$, ${\cal F}$, ${\cal H}$,
${\cal A}$ and $\lambda_{ijk}$. This is mainly because (i) the heavy
moduli $\Phi=\{S, U, \Sigma\}$ stabilized by flux have negligible
$F$-components, $F^{\Phi}/\Phi \sim m_{3/2}^2/m_{\Phi}\ll
m_{3/2}/8\pi^2$, thus do not participate in SUSY-breaking, and (ii)
the SUSY-breaking IR fields $Z$ and $\Lambda^\alpha$ are sequestered
from the observable sector.

The vacuum value of $Z$ is determined by $W_{\rm flux}$, and related
to the metric warp factor $e^{2A}$ at the tip of throat as
\bea Z \sim \exp \Big(-8\pi^2N_{RR}}S_0/{N_{NS}\Big)\sim e^{3A},\eea
where $S_0$ is the vacuum value of $S$ determined by $D_SW=0$. Since
the scalar component of $CC^*$ corresponds to the conformal factor
of $g^{\mu\nu}$, which can be read off from the Weyl transformation
(\ref{weyl}), $C$ in $\Gamma_{IR}$ should appear in the combination
$Ce^A\sim CZ^{1/3}$. Then the $C$-dependence determined by the Weyl
invariance (\ref{weyl}) suggests  that \bea
\Gamma_{IR}^{(0)}&\sim& (ZZ^*)^{1/3}\,\sim\, e^{2A}, \nonumber \\
\Gamma_{IR}^{(1)}&\sim& Z\,\sim\, e^{3A}, \nonumber \\
\Gamma_{IR}^{(2)}&\sim& (ZZ^*)^{2/3} \,\sim\, e^{4A}\eea for which
\bea m_Z\,\sim\, \frac{F^Z}{Z}\,\sim\, e^A \eea as anticipated. Here
and in the following, unless specified, we use the unit with the 4D
Planck scale $M_{Pl}=1/\sqrt{8\pi G_N}=1$.

The SUSY breaking at the tip of throat provides a positive vacuum
energy density of the order of $M_{\rm SUSY}^4\sim e^{4A}$. This
positive vacuum energy density should be cancelled by the negative
SUGRA contribution of the order of $m_{3/2}^2$, which requires \bea
m_{3/2}\sim e^{2A}.\eea One then finds the following pattern of mass
scales \cite{choi1}: \bea && m_{S,U,\Sigma}\sim
\frac{1}{M_{st}^2R^3}\sim
10^{15} \, {\rm GeV}, \nonumber \\
&& m_Z\sim e^A M_{st}\sim 10^{10} \, {\rm GeV}, \nonumber \\
&&m_T\sim m_{3/2}\ln(M_{Pl}/m_{3/2})\sim 10^6\, {\rm GeV}, \nonumber
\\
&&m_{3/2}\sim m_{\rm soft}\ln(M_{Pl}/m_{3/2})\sim 10^4\, {\rm GeV}
 \eea where $m_{\rm soft}$ denotes the soft
masses of the visible fields, e.g. the gaugino masses, and the
string scale $M_{st}$ and the CY radius $R$ are given by $M_{st}\sim
\frac{1}{R}\sim 10^{17}$ GeV.

 The heavy moduli
$S,U,\Sigma$ and the throat modulus $Z$ couple to the light visible
fields and $T$ only through the Planck scale suppressed
interactions. Those hidden sector fields can be integrated out to
derive an effective action of $V^a,Q^i,T$ and the Goldstino
superfield $\Lambda^\alpha$ renormalized at a high scale
near $M_{GUT}$. After this procedure, the effective action can be
written as \cite{choi1,choi2}\bea \label{4daction2}&& \int
d^4x\sqrt{g}\left[ \,\int d^2\theta \,\left( \frac{1}{4}f^{\rm
eff}_a W^{a\alpha}W^a_\alpha+C^3W_{\rm eff}\right) \right.
\nonumber \\
&&\left.\qquad\qquad+\,\int d^4\theta \, CC^*\Omega_{\rm eff}
\right],\eea where
\bea f_a^{\rm eff}&=&T+lS_0,\nonumber \\
 \Omega_{\rm
eff}&=&-3e^{-K_0/3}+{\cal
Y}_iQ^{i*}Q^i-e^{4A}CC^*\Lambda^2\bar{\Lambda}^2{\cal P}_{\rm
lift}\nonumber
\\
&-&\Big(\frac{e^{3A}C^{*2}}{C}\Lambda^2\Gamma_0+{\rm h.c}\Big),
\nonumber
\\
W_{\rm eff}&=&w_0+{\cal
A}e^{-8\pi^2(k_1T+l_1S_0)}+\frac{1}{6}\lambda_{ijk}Q^iQ^jQ^k,\nonumber\eea
where $S_0=\langle S\rangle$, $K_0=K_0(T+T^*)$ is the K\"ahler
potential of $T$, $e^{K_0/3}{\cal Y}_i$ is the  K\"ahler metric of
$Q^i$,
${\cal P}_{\rm lift}$ and $\Gamma_0$ are constants of order unity,
and finally $w_0$ is the vacuum value of $W_{\rm flux}$. Note that
at this stage, all of $e^{2A}, {\cal P}_{\rm lift}, \Gamma_0, S_0,
w_0$, and ${\cal A}$ correspond to field-independent constants
obtained after $S,U,\Sigma$ and $Z$ are integrated out. As we have
noticed, the condition for vanishing cosmological constant requires
\bea w_0\sim e^{2A}\sim e^{-8\pi^2 l_0 S_0} \qquad
\Big(l_0=\frac{2N_{RR}}{3N_{NS}}\Big),\eea and the weak scale SUSY
can be obtained for the warp factor value $e^{2A}\sim 10^{-14}$. For
such a small value of  warp factor, one finds that the SUSY-breaking
$F$ components are determined as follows {\it independently of} the
moduli K\"ahler potential $K_0$ \cite{choi1,choi2,choi3}: \bea
\label{fterms} \frac{F^C}{C}&=&m_{3/2}\left(1+{\cal
O}\left(\frac{1}{4\pi^2}\right)\right),\nonumber \\
\frac{F^T}{T+T^*}&=&\frac{l_0}{l_0-l_1}\frac{m_{3/2}}{\ln(M_{Pl}/m_{3/2})}\left(1+{\cal
O}\left(\frac{1}{4\pi^2}\right)\right), \nonumber \\
F^{S,U,\Sigma}&\sim &\frac{m_{3/2}^2}{m_{S,U,\Sigma}}\,\ll
\,\frac{m_{3/2}}{8\pi^2}.\eea

Basically same pattern of $F$ components is obtained in more general
set-up with arbitrary number of K\"ahler moduli $T_I$. \cite{choi5}.
 Without loss of generality, one
can choose a field basis $T_I=\{T_x,T_\alpha\}$, for which the
superpotential is given by \bea
W_{\rm eff}=w_0+\sum_x{\cal A}_x e^{-8\pi^2(k_xT_x+l_xS_0)}, \eea
where ${\cal A}_x$ are constants of order unity, while $w_0\sim
e^{2A}$ as required to tune the cosmologically constant to be nearly
zero. If the moduli K\"ahler potential admits a solution for
$\partial K_0/\partial T_\alpha=0$ in this field basis, one finds
that $T_x$ are stabilized by nonperturbative terms in $W_{\rm eff}$
with a mass $m_{T_x}\sim m_{3/2}\ln(M_{Pl}/m_{3/2})$, while ${\rm
Re}(T_\alpha)$ are stabilized essentially by the uplifting potential
$V_{\rm lift}=e^{4A}{\cal P}_{\rm lift}e^{2K_0/3}$ with a mass
$m_{{\rm Re}(T_\alpha)}\sim m_{3/2}$. (${\rm Im}(T_\alpha)$ are
nearly massless axions one of which might solve the strong CP
problem.) Still the $F$-components follow the pattern
\cite{choi5}\bea \label{qq}\frac{F^{T_x}}{T_x+T_x^*}&\sim&
\frac{F^{T_\alpha}}{T_\alpha+T_\alpha^*}\sim
\frac{m_{3/2}}{\ln(M_{Pl}/m_{3/2})},\nonumber \\
F^{S,U,\Sigma}&\sim &\frac{m_{3/2}^2}{m_{S,U,\Sigma}}\eea which is
basically same as (\ref{fterms}).

One of the interesting features of SUSY breaking at the IR end of
throat is the {\it sequestering} property,
i.e. there is no sizable Goldstino-matter contact term: \bea \Delta
m_i^2CC^*\Lambda^2\bar{\Lambda}^2Q^{i*}Q^i\eea in $\Omega_{\rm eff}$
of (\ref{4daction2}), which would give an additional contribution
$\Delta m_i^2$ to the soft scalar mass-squares. This amounts to that
there is no operator of the form $(ZZ^*)^{1/3}Q^{i*}Q^i$ or
$(ZZ^*)^{2/3}\Lambda^2\bar{\Lambda}^2Q^{i*}Q^i$ in $e^{-K/3}$ of
(\ref{4daction1}). Since $Q^i$ and $\Lambda^\alpha$ are
geometrically separated by warped throat, such contact term can be
generated only by the exchange of bulk field propagating through the
throat. Simple operator analysis assures that the exchange of chiral
multiplet can induce only a higher order operator in the superspace
derivative expansion, while the exchange of light vector multiplet
$\tilde{V}$ can generate the Goldstino-matter contact term with
$\Delta m_i^2\sim \langle D_{\tilde{V}}\rangle$, where
$D_{\tilde{V}}$ is the $D$-component of $\tilde{V}$
\cite{choi3,hebecker2}. Quite often, throat has an isometry symmetry
providing light vector field which might generate the
Goldtino-matter contact term. However, in many cases, the isometry
vector multiplet does not develop a nonzero $D$-component, and
thereby not generate the contact term \cite{choi3,kachrusundrum}. As
an example, let us consider the SUSY breaking by anti-$D3$ brane
stabilized at the tip of Klebanov-Strassler (KS)  throat which has
an $SO(4)$ isometry \cite{ks}. Adding anti-$D3$ at the tip breaks
SUSY and also $SO(4)$ down to $SO(3)$. However the unbroken $SO(3)$
assures that the $SO(4)$ vector multiplets have vanishing
$D$-components, thus do not induce the Goldstino-matter contact
term. In fact, this is correct only up to ignoring the
isometry-breaking deformation of KS throat, which is caused by
attaching the throat to  compact CY. Recently, the effect of such
deformation has been estimated \cite{kachrusundrum}, which found
\bea \Delta m_i^2\,\leq\, {\cal O}(e^{\sqrt{28}A})\,\sim\,
10^{-8}m_{3/2}^2.\eea This is small enough to be ignored compared to
the effects of $F^C$ and $F^T$ obtained in (\ref{fterms}).

\section{mirage mediation pattern of sparticle masses}

The results (\ref{fterms}) and (\ref{qq})  on SUSY-breaking
$F$-components indicates that \bea F^T/T\sim \frac{m_{3/2}}{4\pi^2}
\gg |F^{S,U,\Sigma}|,\eea where $T$ denotes generic K\"ahler moduli.
In such case, soft terms are determined dominantly by the K\"ahler
moduli-mediated contribution and the one-loop anomaly mediated
contribution which are comparable to each other.
 For the canonically normalized soft terms: \bea
 -\frac{1}{2}M_a\lambda^a\lambda^a-\frac{1}{2}m_i^2|\phi^i|^2
-\frac{1}{6}A_{ijk}y_{ijk}\phi^i\phi^j\phi^k+{\rm h.c.}, \eea where
$\lambda^a$ are gauginos, $\phi^i$ are sfermions, $y_{ijk}$ are the
canonically normalized Yukawa couplings, the soft parameters at
energy scale just below  $M_{GUT}$ are given by \bea \label{soft1}
M_a &=& M_0 +\frac{b_a}{16\pi^2}g_{GUT}^2m_{3/2},
\nonumber \\
A_{ijk} &=&
\tilde{A}_{ijk}-\frac{1}{16\pi^2}(\gamma_i+\gamma_j+\gamma_k)m_{3/2},
\nonumber \\
m_i^2 &=& \tilde{m}_i^2-\,\frac{1}{32\pi^2}\frac{d\gamma_i}{d\ln
\mu}m_{3/2}^2 \nonumber \\
&+&\frac{1}{4\pi^2}
\left[\sum_{jk}\frac{1}{4}|y_{ijk}|^2\tilde{A}_{ijk} -\sum_a
g^2_aC^a_2(\phi^i)M_0\right]m_{3/2}, \nonumber \eea where the
moduli-mediated soft masses $M_0$, $\tilde{A}_{ijk}$ and
$\tilde{m}_i^2$ are given
by \bea M_0 &=& F^T\partial_T\ln({\rm Re}(f_a)) \nonumber \\
 \tilde{A}_{ijk}&=& F^T\partial_T\ln({\cal Y}_i{\cal Y}_j{\cal Y}_k), \nonumber \\
\tilde{m}^2_i&=& -|F^T|^2\partial_T\partial_{\bar{T}}\ln({\cal
Y}_i), \eea  and $b_a = -3{\rm tr}\left(T_a^2({\rm
Adj})\right)+\sum_i {\rm tr}\left(T^2_a(\phi^i)\right)$, $\gamma_i
=2 \sum_a g_a^2C_2^a(\phi^i)-\frac{1}{2}\sum_{jk}|y_{ijk}|^2$, where
$C_2^a(\phi^i)=(N^2-1)/2N$ for a fundamental representation $\phi^i$
of the gauge group $SU(N)$, $C_2^a(\phi^i)=q_i^2$ for the $U(1)$
charge $q_i$ of $\phi^i$, and
$\omega_{ij}=\sum_{kl}y_{ikl}y^*_{jkl}$ is assumed to be diagonal.

Taking into account the 1-loop RG evolution, the above soft masses
at $M_{GUT}$ lead to the following low energy gaugino masses \bea
\label{ppp} M_a(\mu) = M_0 \left[\,1-\frac{1}{8\pi^2}b_ag_a^2(\mu)
\ln\left(\frac{M_{\rm mir}}{\mu}\right)\,\right], \eea showing that
the gaugino masses are unified at the {\it mirage messsenger scale}
\cite{choi2}: \bea \label{mirage-scale} M_{\rm mir} =
\frac{M_{GUT}}{(M_{Pl}/m_{3/2})^{\alpha/2}}, \eea where \bea
\label{alpha} \alpha \equiv \frac{m_{3/2}}{M_0\ln(M_{Pl}/m_{3/2})},
\nonumber\eea
 while the gauge
couplings are still unified at $M_{GUT}=2\times 10^{16}$ GeV. With
this feature of mirage unification,   the SUSY breaking scheme
discussed above has been named as mirage mediation \cite{nilles1}.
The low energy values of $A_{ijk}$ and $m_i^2$ generically depend on
the associated Yukawa couplings $y_{ijk}$. However if $y_{ijk}$ are
negligible or if
$\tilde{A}_{ijk}/M_0=(\tilde{m}^2_i+\tilde{m}^2_j+\tilde{m}^2_k)/M_0^2=1$,
their low energy values also show  the mirage unification feature
\cite{choi2}:\bea \label{mirage-solution} &&A_{ijk}(\mu) =
\tilde{A}_{ijk}+ \frac{M_0}{8\pi^2}(\gamma_i + \gamma_j+\gamma_k)
\ln\left(\frac{M_{\rm mir}}{\mu}\right),
\nonumber \\
&&m_i^2(\mu) = \tilde{m}_i^2-\frac{M^2_0}{8\pi^2}Y_i\left(
\sum_jc_jY_j\right)g^2_Y\ln\left(\frac{M_{GUT}}{\mu}\right)
\nonumber \\
&+&\frac{M_0^2}{4\pi^2}\left\{
\gamma_i-\frac{1}{2}\frac{d\gamma_i}{d\ln \mu}\ln\left( \frac{M_{\rm
mir}}{\mu}\right)\right\} \ln\left( \frac{M_{\rm mir}}{\mu}\right),
\eea where $Y_i$ is the $U(1)_Y$ charge of $\phi^i$. Quite often,
the moduli-mediated squark and slepton masses have a common value,
i.e. $\tilde{m}^2_{\tilde{Q}}=\tilde{m}^2_{\tilde{L}}$, and then the
 squark and slepton masses of the  1st and 2nd generation are unified again
at $M_{\rm mir}$.

Mirage mediation can be generalized in a way which includes a
comparable size of gauge mediation \cite{axionic,deflected,choi4}.
As an example, one can consider a model with exotic vector-like
matter fields $\Phi+\Phi^c$ described by \bea \label{deflectedm}
&&\int d^4\theta CC^*\Big({\cal Y}_\Phi\Phi^*\Phi+ {\cal
Y}_{\Phi^c}\Phi^{c*}\Phi^c\Big) \nonumber \\
&&+\,\int d^2\theta C^3\Big(\kappa X^n+\lambda X\Phi^c\Phi\Big)
\quad (n>3),\eea where $\kappa\sim 1/M_{Pl}^{n-3}$. One then finds
\cite{deflected}\bea F^X/X &=&-\frac{2}{n-1}\frac{F^C}{C},
\nonumber \\
\langle X\rangle &\sim&
\left(m_{3/2}M_{Pl}^{n-3}\right)^{1/(n-2)}.\eea Integrating out the
massive $\Phi+\Phi^c$  gives rise to the following gauge threshold
correction to soft parameters at the gauge messenger scale $\langle
X\rangle$: \bea \Delta M_a(\langle
X\rangle)&=&-\frac{N_\Phi}{16\pi^2}g_a^2\left(\frac{F^X}{X}+\frac{F^C}{C}\right),
\nonumber \\
\Delta m_i^2(\langle X\rangle)&=&\frac{2N_\Phi}{(16\pi^2)^2}\sum_a
g_a^4C_2^a(\phi^i)\left|\frac{F^X}{X}+\frac{F^C}{C}\right|^2, \nonumber \\
\Delta A_{ijk}(\langle X\rangle)&=&0,\eea where $N_\Phi$ denotes the
number of $\Phi+\Phi^c$ which is assumed to be $5+\bar{5}$ of
$SU(5)$. These gauge mediation contributions are  comparable to the
moduli and anomaly mediations, and alter the shape of low energy
sparticle spectra.

For the gauginos and matter families with small Yukawa coupling and
$\sum_ic_iY_i=0$, one can find the approximate analytic expression
of low energy sparticle masses at $\mu< \langle X\rangle$
\cite{choi4}: \bea M_a(\mu) &=& M^{\rm eff}_0
\left[\,1-\frac{1}{8\pi^2}b_ag_a^2(\mu) \ln\left(\frac{M^{\rm
eff}_{\rm mir}}{\mu}\right)\,\right], \nonumber \\
 m_i^2(\mu) &=&
\left(\tilde{m}^{{\rm eff}}_i\right)^2+\left\{
\gamma_i-\frac{1}{2}\frac{d\gamma_i}{d\ln \mu}\ln\left( \frac{M^{\rm
eff}_{\rm mir}}{\mu}\right)\right\}\nonumber \\
&&\times \ln\left( \frac{M^{\rm eff}_{\rm
mir}}{\mu}\right)\frac{\left(M^{{\rm eff}}_0\right)^2}{4\pi^2},
 \eea
 where
 \bea
M_0^{\rm
 eff}&=&RM_0,\nonumber \\
M_{\rm mir}^{\rm eff}&=&\frac{M_{GUT}}{(M_{Pl}/m_{3/2})^{\alpha/2R}}
 \nonumber \\
(\tilde{m}_i^{\rm eff})^2&=&\tilde{m}_i^2+
 \left[\frac{2}{N_\Phi}\sum_aC_2^a(\phi^i)\frac{g_a^4(\langle
 X\rangle)}{g_0^4}
 \right.
 \nonumber\\
 &-&\left.\sum_a\frac{2C_2^a(\phi^i)}{b_a}\left(1-\frac{g_a^4(\langle X\rangle)}{g_0^4}\right)\right]
 \left(1-R\right)^2M_0^2
 \nonumber \\
 &+&\,\left[4\sum_a\frac{C_2^a(\phi^i)}{b_a}\left(1-\frac{g_a^2(\langle
 X\rangle)}{g_0^2}\right)\right](1-R)M_0^2\nonumber
 \eea
 for\bea
 M_0&=&F^T\partial_T\ln({\rm Re}(f_a)),\nonumber \\
R&=& \left[1-\frac{N_\Phi
g_0^2}{8\pi^2}\ln\left[\left(\frac{M_{Pl}}{m_{3/2}}\right)^{\alpha/2\beta}\frac{M_{GUT}}{\langle
 X\rangle}\right]\right],\nonumber \\
 \alpha&=&\frac{m_{3/2}}{M_0\ln(M_{Pl}/m_{3/2})},\nonumber \\
 \beta&=&\frac{m_{3/2}}{F^X/X}.
 \eea
Here
 $b_a$ are the one-loop beta function coefficients at a scale
 $\mu<\langle X\rangle$
 and $g_0^2\simeq 1/2$ corresponds to the unified
 gauge coupling constant in the absence of exotic matter fields
 $\Phi+\Phi^c$.
Note that the gaugino masses are still unified at a mirage scale
$M_{\rm mir}^{\rm eff}$ even when there exists a sizable extra gauge
mediation contribution, while the mirage unification of sfermion
masses is generically lost. Although the mirage unification of
sfermion masses is generically lost in the presence of gauge
mediation, the deviation is not so significant for the class of
models giving $\beta<0$ \cite{choi4}. For instance, for the models
of (\ref{deflectedm}),  $|R-1|={\cal O}(0.1)$ for a reasonable range
of $\alpha, N_\Phi$ and $n>3$, indicating that sfermion masses show
a mirage unification at the same scale as the gaugino masses up to
small deviations of ${\cal O}(10)$ \%.

In regard to phenomenology, the most interesting feature of mirage
mediation is that it gives rise to {\it significantly compressed low
energy SUSY spectrum} compared to other popular schemes such as
mSUGRA, gauge mediation and anomaly mediation. This feature can be
easily understood by noting that soft parameters are unified at
$M_{\rm mir}=M_{GUT}(m_{3/2}/M_{Pl})^{\alpha/2}$ which is
hierarchically lower than $M_{GUT}$ as $\alpha$ has  a positive
value of order unity.

In fact, mirage mediation provides more concrete prediction under a
rather plausible assumption. In the following, we present some
predictions of the minimal mirage mediation yielding the low energy
soft parameters given by (\ref{ppp}) and (\ref{mirage-solution}).
 Assuming that $f_a$ are (approximately) universal, which might be
required to realize the gauge coupling unification at $M_{GUT}$, the
low energy gaugino masses at TeV are given by \cite{choinilles}
 \bea M_1&\simeq & M_0(0.42+0.28\alpha),\nonumber \\
M_2&\simeq & M_0(0.83+0.085\alpha),\nonumber \\
M_3&\simeq& M_0(2.5-0.76\alpha),\eea leading to
\bea && M_1:M_2:M_3\nonumber \\
&\simeq& (1+0.66\alpha):(2+0.2\alpha):(6-1.8\alpha).\eea
 The TeV scale masses of
the 1st and 2nd generations of squarks and sleptons are also easily
obtained to be \bea m_{\tilde{Q}}^2&\simeq&
\tilde{m}^2_{\tilde{Q}}+M_0^2(5.0-3.6\alpha+0.51\alpha^2),\nonumber
\\
m_{\tilde{D}}^2&\simeq&\tilde{m}^2_{\tilde{D}}+M_0^2(4.5-3.3\alpha+0.52\alpha^2),\nonumber
\\
m_{\tilde{L}}^2&\simeq&\tilde{m}^2_{\tilde{L}}+M_0^2(0.49-0.23\alpha-0.015\alpha^2),\nonumber
\\
m_{\tilde{E}}^2&\simeq&\tilde{m}^2_{\tilde{E}}+M_0^2(0.15-0.046\alpha-0.016\alpha^2),
\eea where $\tilde{Q},\tilde{D},\tilde{L}$ and $\tilde{E}$ denote
the $SU(2)_L$ doublet squark,  singlet down-squark, doublet lepton,
and singlet lepton, respectively. Assuming that the matter K\"ahler
metrics obey simple unification (or universality) relations such as
${\cal Y}_Q={\cal Y}_E$ and ${\cal Y}_D={\cal Y}_L$, we find \bea
&&M_1^2:(m_{\tilde{Q}}^2-m_{\tilde{E}}^2):(m_{\tilde{D}}^2-m_{\tilde{L}}^2)
\nonumber \\
&\simeq& (0.18+0.24\alpha+0.09\alpha^2): \nonumber
\\
&&(4.9-3.5\alpha+0.53\alpha^2):\nonumber\\
&& (4.0-3.1\alpha+0.54\alpha^2).\eea

If the idea of low energy SUSY is correct and the gluino or squark
masses are lighter than 2 TeV,  some superparticle masses, e.g. the
gluino mass and the first two neutralino masses as well as some of
the squark and slepton masses, might be determined at the LHC by
analyzing various kinematic invariants of the cascade decays of
gluinos and squarks. It is then quite probable that the LHC
measurements of those superparticle masses are good enough to test
the above predictions of mirage mediation \cite{cho}.


\begin{theacknowledgments}
This work is supported by the KRF Grant funded by the Korean
Government (KRF-2005-201-C00006). I thank  K. S. Jeong, H. P.
Nilles, and K. Okumura for useful discussions.

\end{theacknowledgments}






\end{document}